\documentclass{article}

\PassOptionsToPackage{numbers, compress}{natbib}
\usepackage[preprint]{neurips_2025}
\usepackage[utf8]{inputenc} % allow utf-8 input
\usepackage[T1]{fontenc}    % use 8-bit T1 fonts
\usepackage{hyperref}       % hyperlinks
\usepackage{url}            % simple URL typesetting
\usepackage{booktabs}       % professional-quality tables
\usepackage{amsfonts}       % blackboard math symbols
\usepackage{nicefrac}       % compact symbols for 1/2, etc.
\usepackage{microtype}      % microtypography
\usepackage{xcolor}         % colors
\usepackage{wrapfig}
\usepackage{adjustbox}
\usepackage{array} 
\usepackage{graphicx}
\usepackage{amsmath}
\usepackage{subcaption}
\usepackage{multirow}
\usepackage{amsmath}
\usepackage{amssymb}
\usepackage{mathtools}
\usepackage{amsthm}

\usepackage[utf8]{inputenc} % allow utf-8 input
\usepackage[T1]{fontenc}    % use 8-bit T1 fonts
\usepackage{hyperref}       % hyperlinks
\usepackage{url}            % simple URL typesetting
\usepackage{booktabs}       % professional-quality tables
\usepackage{amsfonts}       % blackboard math symbols
\usepackage{nicefrac}       % compact symbols for 1/2, etc.
\usepackage{microtype}      % microtypography
\usepackage{xcolor}         % colors

\theoremstyle{plain}

\theoremstyle{definition}

\theoremstyle{remark}

\usepackage{float}

\title{\textsc{FinAudio}: A Benchmark for Audio Large Language Models in Financial Applications}

\author{
Yupeng Cao$^{1}$,
Haohang Li$^{1}$,
Yangyang Yu$^{1}$,
Shashidhar Reddy Javaji$^{1}$\\
\textbf{Yueru He$^{2}$,
Jimin Huang$^{3}$,
Qianqian Xie$^{3}$,
Fabrizio Dimino$^{4}$,
Xiao-Yang Liu$^{2}$}\\
\textbf{K.P. Subbalakshmi$^{1}$,
Meikang Qiu$^{5}$,
Sophia Ananiadou$^{6}$,
Jian-Yun Nie$^{7}$}\\
\\[-2mm]
$^{1}$Stevens Institute of Technology,
$^{2}$Columbia University, 
$^{3}$The Fin AI, 
$^{4}$Domyn, \\
$^{5}$Augusta University,
$^{6}$The University of Manchester,
$^{7}$University of Montreal
}

\begin{document}

\maketitle

\begin{abstract}
  Audio Large Language Models (AudioLLMs) have received widespread attention and have significantly improved performance on audio tasks such as conversation, audio understanding, and automatic speech recognition (ASR). Despite these advancements, there is an absence of a benchmark for assessing AudioLLMs in financial scenarios, where audio data, such as earnings conference calls and CEO speeches, are crucial resources for financial analysis and investment decisions. In this paper, we introduce \textsc{FinAudio}, the first benchmark designed to evaluate the capacity of AudioLLMs in the financial domain. We first define three tasks based on the unique characteristics of the financial domain: 1) ASR for short financial audio, 2) ASR for long financial audio, and 3) summarization of long financial audio. Then, we curate two short and two long audio datasets, respectively, and develop a novel dataset for financial audio summarization, comprising the \textsc{FinAudio} benchmark. We evaluate seven prevalent AudioLLMs on \textsc{FinAudio}. Our evaluation reveals the limitations of existing AudioLLMs in the financial domain and offers insights for improving AudioLLMs (\href{https://finaudio-doc.readthedocs.io/en/latest/}{Project Page}).  
\end{abstract}

\section{Introduction}
\label{sec:intro}

Timely and accurate interpretation of financial audio underpins sentiment analysis and investment decision-making~\cite{qin2019you, yang2020html, cao2024ecc}. Financial audio data serves as a critical foundation across diverse scenarios, including earnings conference calls, investor presentations, and customer service interactions. Recent developments in financial AI have been propelled by breakthroughs in language-centric models, particularly large language models (e.g., BloombergGPT~\cite{wu2023bloomberggpt}, PIXIU~\cite{xie2023pixiu}, FinGPT~\cite{liu2023fingpt}) and by evaluation suites such as FinBen~\cite{xie2025finben} that benchmark LLM performance on financial NLP tasks. However, a critical gap remains: none of these models or benchmarks addresses financial audio, and even the latest multimodal model (FinLLaVA-8B~\cite{xie2024open}, designed to handle text, images, and charts) cannot process audio inputs.

In the general domain, AudioLLMs incorporate an audio encoder into language models~\cite{wu2023multimodal, zhang2024mm, radford2023robust, hu2024wavllm, chu2023qwen, chu2024qwen2, tangsalmonn}. This integration allows LLMs to accept audio inputs and perform tasks such as automatic speech recognition (ASR) and spoken question answering (SQA). These audio-capable LLMs have demonstrated effectiveness on general benchmarks~\cite{yang2024air, wang2024audiobench}. Yet, the financial domain still lacks a dedicated audio benchmark, leaving researchers unable to systematically evaluate model performance on financial audio or compare different approaches, thereby hindering progress in audio-driven financial applications, such as investment advisory services.

To address this critical need, we introduce \textsc{FinAudio}, the first AudioLLM benchmark tailored for the financial domain. Specifically, \textsc{FinAudio} includes three tasks representing real-world financial scenarios: (1) ASR for short financial audio clips (each under 1 minute), (2) ASR for long financial audio recordings (45–60 minutes, e.g. earnings calls), and (3) summarization of financial audio data.
To support these tasks, we curated five financial audio datasets. Four are from existing open-source datasets (two with short clips and two with full-length recordings). Additionally, we created a new dataset tailored for financial audio summarization, named FinAudioSum. Altogether, the \textsc{FinAudio} benchmark provides 400+ hours of financial audio data, comparable in scale to general-domain AudioLLM benchmarks.

On \textsc{FinAudio}, we evaluate seven representative AudioLLMs and observed the following: (a) AudioLLMs achieve higher accuracy on short clips than on long recordings; (b) Errors in long-audio ASR directly undermine the quality of audio summaries. (c) AudioLLMs exhibit significant variability in their robustness to different instructions; (d) The open-source model Whisper-v3 outperforms closed-source models on all ASR tasks, offering a low-cost, privacy-preserving solution for financial applications. Our main contributions are below: (1) We present \textsc{FinAudio}, the first comprehensive open-source benchmark for evaluating AudioLLMs on financial audio tasks. (2) We design three realistic financial speech tasks and compile five corresponding datasets, including the new FinAudioSum dataset for audio summarization. (3) We conduct an extensive evaluation of seven AudioLLMs on these tasks, highlighting each model’s strengths and limitations and outlining challenges for future research. \textsc{FinAudio} aims to accelerate the development of robust AudioLLMs for financial applications, promoting transparency and accessibility of financial information processing.
\vspace{-1em}

%To further this research area, we provide an in-depth discussion of the challenges and outline promising directions for future research in financial audio applications. 
\section{The \textsc{FinAudio} Benchmark}
\label{sec:finaudio}
\begin{figure}[h]
\begin{center}
\centerline{\includegraphics[width=0.80\columnwidth]{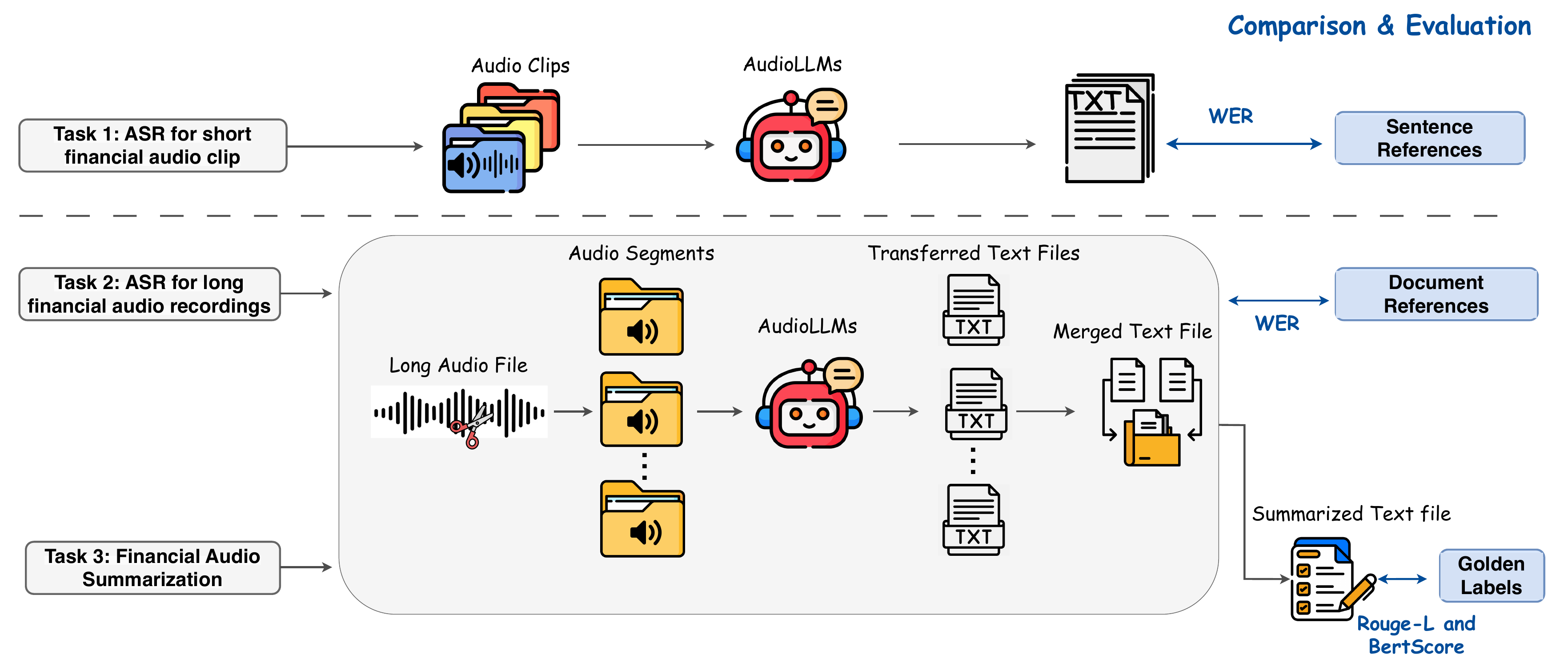}}
\caption{Evaluation pipelines for the three \textsc{FinAudio}'s tasks.}% \textit{Note: WER refers to the term of Word Error Rate.}}
\label{fig:task}
\end{center}
\vspace{-3em}
\end{figure}

\vspace{-0.5em}
Figure~\ref{fig:task} is an overview of \textsc{FinAudio}. This section defines three key tasks in FinAudio and describes how we constructed the evaluation datasets. Each task corresponds to a specific real-world financial audio scenario, allowing us to test different aspects of an AudioLLM’s capabilities. For each task, we both compile existing datasets and develop new ones to ensure a comprehensive evaluation. 

\subsection{\textsc{FinAudio} Tasks}
\vspace{-0.5em}
% \textsc{FinAudio}  aims to evaluate the performance of AudioLLMs on financial audio data. Our evaluation primarily focuses on the models' ability to comprehend financial audio across various financial scenarios. Accordingly, our benchmarks are designed to evaluate three crucial aspects:

\subsubsection{\textbf{ASR for short financial audio clip}}
\vspace{-0.5em}
Automatic Speech Recognition converts spoken language into text. Short audio clips from financial chatbot assistants or financial news often contain dense financial terms. This task evaluates AudioLLMs' accuracy in transcribing financial audio clips, directly impacting applications such as financial voice assistants. In this task, each input is represented as $(A, Q, R)$, where $A$ is an audio clip, $Q$ is the prompt instruction, and $R$ is the reference transcript. The transcribed text $T$ is generated as $T = \textit{AudioLLM}(A, Q)$. The Word Error Rate (WER) is computed as: $\text{WER} = \frac{S + D + I}{N}$, where $S$, $D$, and $I$ represent the number of substitutions, deletions, and insertions, respectively, and $N$ is the total number of words in the reference ($N = S + D + C$, with $C$ indicating correct words). A lower WER indicates better ASR performance.

% \begin{equation}
%     \text{WER} = \frac{S + D + I}{N}
% \label{eq:wer}
% \end{equation}

% In this task, each input data is represented as $(A, Q, R)$, where $A$ is the audio clip, $Q$ denotes the prompt instruction, and $R$ is a reference. The transcribed text $T$ is generated by $T = \textit{AudioLLM(A, Q)}$. We then compute the Word Error Rate (WER) between $T$ and $R$ using the equation: $WER = \frac{S + D + I}{N}$, where $S$ is the number of substitutions, $D$ is the number of deletions, $I$ is the number of insertions, $C$ is the number of correct words, and $N$ is the number of words in the reference (N=S+D+C). A lower WER indicates a more robust ASR capability of the model.

\subsubsection{\textbf{ASR for long financial audio recordings}}
\label{long_asr}
Long audio recordings are common in finance, capturing events like quarterly earnings calls, mergers and acquisitions discussions, and central bank meetings. In this scenario, ASR task is thus used to transcribe lengthy financial audio, evaluating AudioLLMs' capability to handle extended audio inputs. Challenges include long durations and complex financial terminology. This task assesses the models' ability to consistently maintain transcription accuracy over long recordings, which is essential for comprehensive financial analysis. To handle the varying window lengths and the inability of current AudioLLMs to process long audio files within a single window, we standardize audio inputs to 30-second segments, dividing each raw audio $A$ into chunks $\{a_1, a_2, \dots, a_n\}$. We transcribe each chunk $a_i$ using AudioLLMs into $t_i$, and concatenated the resulting transcriptions as $T = [t_1; t_2; \dots; t_n]$. WER then evaluates the transcription quality.

% Due to the varying window lengths of existing AudioLLMs and none that can process the entirety of a long audio file within a single window, we standardized the input audio length to 30 seconds to maintain uniformity in the evaluation process. Consequently, raw audio $A$ is segmented into a series of chunks $\{a_1, a_2, ..., a_n\}$. Each chunk $a_i$ with a maximum 30-second length is then processed by the AudioLLMs to produce its corresponding transcription $t_i$. These transcriptions are subsequently concatenated to form $T = [t_1; t_2; ...; t_n]$. we compute the WER using Equation 1 to evaluate the transcriptions.

\subsubsection{\textbf{Financial audio summarization}}
This task assesses the ability of AudioLLMs to accurately summarize lengthy financial audio content, a scenario specific to financial applications. It evaluates models' understanding of financial discussions, extraction of key points, and effectiveness in condensing lengthy audio into concise summaries. This capability is crucial for stakeholders needing quick insights from extensive financial dialogues, such as investor briefings or regulatory meetings.

AudioLLMs can process audio with a maximum length of 30 seconds. Therefore, AudioLLM cannot perform the summarization task independently, we developed a processing pipeline. Given input data represented as $(A, Q, L)$, where $A$ denotes the long audio recording, $Q$ the prompt instruction, and $L$ the reference summary, the pipeline first segments the long audio into smaller chunks (see Section~\ref{long_asr}). Each chunk is transcribed by an AudioLLM into text segments $t_i$, which are combined into a complete transcription $T$. This transcription is then input into an LLM to generate the summary $S$. Summarization performance is evaluated using two widely adopted summarization evaluation metrics: Rouge-L~\cite{lin2004rouge} and BertScore~\cite{zhangbertscore}, computed between the generated summary ($S$) and the reference summary ($L$). The higher values for these metrics indicate better AudioLLM performance.

% \begin{equation}
%     \text{Rouge-L} = \frac{(1+\beta^2) \cdot R_{\text{LCS}} \cdot P_{\text{LCS}}}{R_{\text{LCS}} + \beta^2 \cdot P_{\text{LCS}}}
% \label{eq:rough-l}
% \end{equation}

% \noindent where \(\beta\) is a weighting factor. \(\text{LCS}(X,Y)\) denotes the length of the longest common subsequence between the candidate summary \(X\) and the reference summary \(Y\). \(|X|\) and \(|Y|\) are the lengths of the candidate and reference summaries, respectively. \(R_{\text{LCS}}\) (Recall) and \(P_{\text{LCS}}\) (Precision) are defined by:
%     \[
%     R_{\text{LCS}} = \frac{\text{LCS}(X,Y)}{|Y|}, \quad
%     P_{\text{LCS}} = \frac{\text{LCS}(X,Y)}{|X|}
%     \]

% \begin{equation}
%     \text{BertScore}(X, Y) = \frac{1}{|X|}\sum_{x_i \in X} \max_{y_j \in Y} x_i^\top y_j
% \label{eq:bertscore}
% \end{equation}

% \noindent where \(x_i\)is the contextual embedding vector of the \(i\)-th token in the candidate summary \(X\). \(y_j\) is the contextual embedding vector of the \(j\)-th token in the reference summary \(Y\). \(x_i^\top y_j\) is the cosine similarity between embeddings \(x_i\) and \(y_j\).

\subsection{\textsc{FinAudio} Datasets}
\label{sec:data}
% The \textsc{FinAudio} benchmark draws from both existing open-source datasets and a newly introduced dataset, covering short financial clips, long earnings calls, and summarization tasks. As summarized in Table~\ref{tab:dataset}, it includes MDRM-test and SPGISpeech-test for short ASR, Earnings-21 and Earnings-22 for long ASR, and our newly developed FinAudioSum for audio summarization. Together, these five datasets provide over 430 hours of diverse financial audio. Detailed dataset descriptions are provided in Appendix~\ref{sec:dataset}.

The FINAUDIO benchmark contributes a unified framework by careful curation and task reformulation, transforming existing datasets into meaningful and comprehensive evaluation tasks. As summarized in Table~\ref{tab:dataset}, we construct MDRM-test and SPGISpeech-test by further developing their respective existing datasets for short ASR evaluation, create Earnings-21 and Earnings-22 test sets for long ASR tasks, and introduce our newly structured FinAudioSum for audio summarization task. Through this curation process, we establish over 430 hours of purposefully organized financial audio that addresses domain-specific evaluation needs not met by the original datasets. Detailed dataset descriptions are provided in Appendix~\ref{sec:dataset}.

\begin{table*}[h]
\centering
\resizebox{0.9\linewidth}{!}{
\begin{tabular}{l|c|r|r|c|c}
\toprule
Dataset Name & Type        & \#Samples & \# Hours & Task                               & Metrics              \\ \midrule
MDRM-test         & Short Clips & 22,208     & 87       & short financial clip ASR           & WER                  \\
SPGISpeech-test   & Short Clips & 39,341     & 130      & short financial clip ASR           & WER                  \\
Earning-21   & Long Audio  & 44        & 39       & long financial audio ASR           & WER                  \\
Earning-22   & Long Audio  & 125       & 120      & long financial audio ASR           & WER                  \\
FinAudioSum  & Long Audio  & 64        & 55       & long financial audio Summarization & Rouge-L \& BertScore \\ \bottomrule
\end{tabular}}
\caption{Statistics of the datasets in the FinAudio benchmark.}
\label{tab:dataset}
\vspace{-1em}
\end{table*}
\section{Experiment Results and Analysis}
We describe the evaluated AudioLLMs and the experiment setup in the appendix~\ref{setup}.

\subsection{Results on ASR Tasks}
\paragraph{Results on short financial audio clips ASR.}  Table~\ref{tab:comparison} shows that Whisper-v3 achieves notably low WER scores (2\%–3\%) on both datasets, highlighting its robust speech recognition without domain-specific tuning. The Qwen2-Audio series (base and instruct versions), GTP-4o-audio, and Gemini exhibit moderate performance with WER scores between 4\% and 6\%. Conversely, SALMONN (7B and 13B) demonstrates substantially lower effectiveness, with WER around or exceeding 40\%–50\%. Despite SALMONN's limited performance, these findings highlight the potential of other mainstream AudioLLMs to support voice-based financial services, particularly conversational assistants handling iterative short audio communications.

\begin{table*}[ht]
\centering
\renewcommand{\arraystretch}{1.2}
\begin{adjustbox}{max width=0.95\textwidth}
\begin{tabular}{lc|c|c|c|c|c|c|c}  % <-- added one more column (c)
\toprule
\multirow{3}{*}{\textbf{Datasets}} & \multicolumn{8}{c}{\textbf{Models}} \\   
\cline{2-9}
 & \textbf{Whisper} & \textbf{Qwen2-Audio} & \textbf{Qwen2-Audio} & \textbf{SALMONN} & \textbf{SALMONN} & \textbf{Gemini} & \textbf{Gemini} & \textbf{GPT-4o-audio} \\ % <-- added header
 & \textbf{-v3} & \textbf{-7B} & \textbf{-7B-Instruct} & \textbf{-7B} & \textbf{-13B} & \textbf{-1.5-flash} & \textbf{-2.0-flash } & \textbf{-transcribe} \\  % <-- subheader
\midrule
MDRM-test & \textbf{2.14} & 3.97 & 4.68 & 51.52 & 49.17 & 4.850 & 4.321 & 4.23 \\   % <-- new col placeholder
SPGISpeech-test & \textbf{2.88} & 4.42 & 5.74 & 39.51 & 41.17 & 5.802 & 5.143 & 4.66\\
Earning-21 & \textbf{11.85} & 26.06 & 29.58 & 83.20 & 80.54 & 18.58 & 19.17 & 15.78 \\
Earning-22 & \textbf{15.93} & 42.76 & 33.65 & 88.50 & 86.23 & 27.13 & 28.12 & 21.37 \\
\bottomrule
\end{tabular}
\end{adjustbox}
\caption{Comparison of model performance based on WER across various speech datasets on ASR.}
\label{tab:comparison}
\vspace{-1em}
\end{table*}

% \begin{figure}[t]
% \begin{center}
% \centerline{\includegraphics[width=0.9\columnwidth]{img/result.png}}
% \caption{WER on models across different datasets. Whisper-v3 consistently achieved the lowest WER.}
% \label{fig:results}
% \end{center}
% \vspace{-2em}
% \end{figure}

\paragraph{Results on long financial audio recording ASR.}
\begin{wrapfigure}{r}{.45\textwidth}
    \centering
    \includegraphics[width=\linewidth]{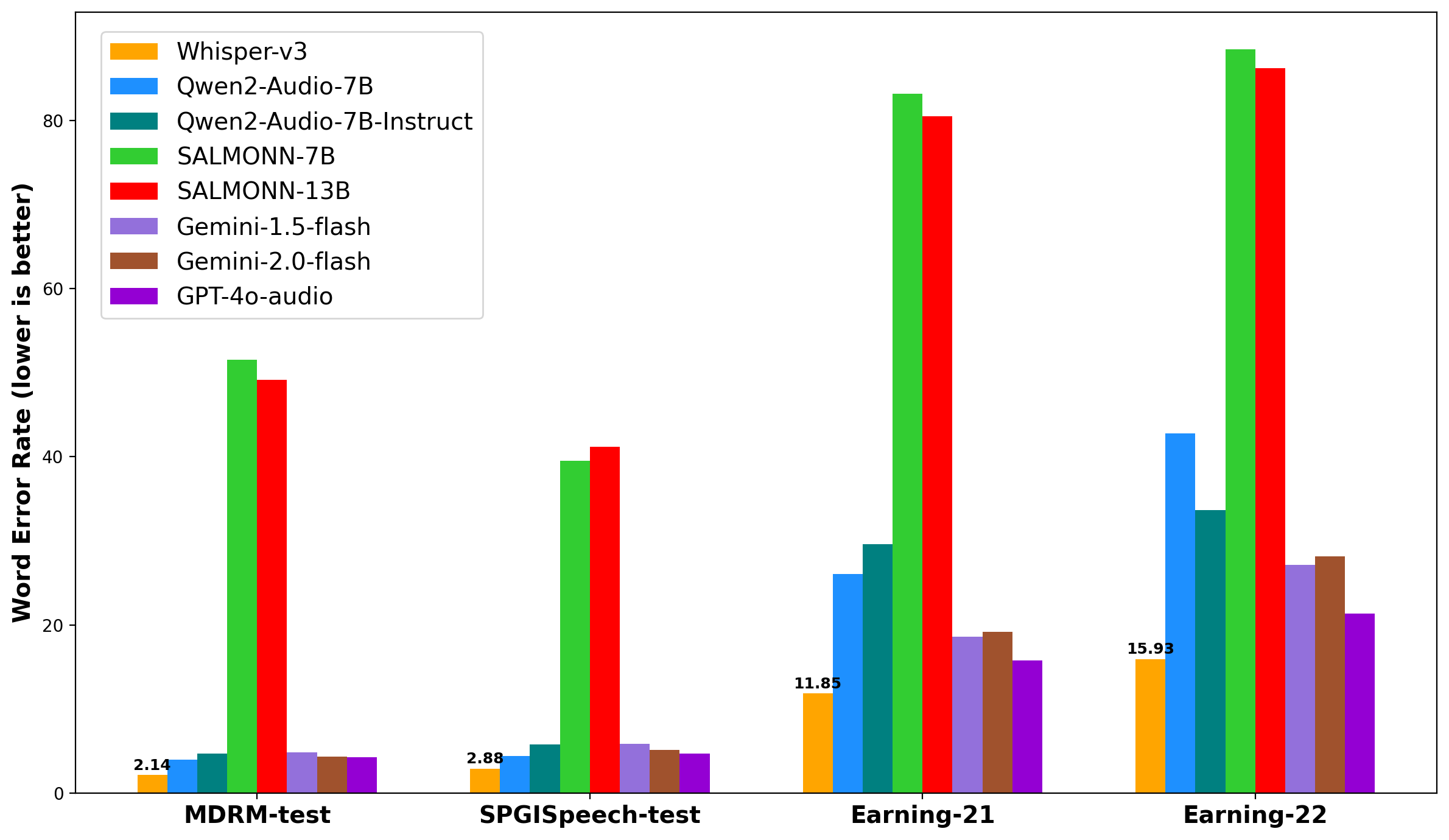}
    \caption{WER on models across different datasets. Whisper-v3 consistently achieved the lowest WER.}
    \label{fig:results}
\end{wrapfigure}
In the long financial audio ASR task, Whisper-v3 remained the top performance. It achieves lower WER in the range of approximately 12\% - 16\%. The GPT-4o-audio emerged as the second-best performer, with WER spanning roughly 15\%–21\%. The error rate of Qwen2-Audio series models has also increased significantly. SALMONN once again reported the highest error rates across both financial datasets (80–88\%). These results reveal a decline in performance across all AudioLLMs, indicating that AudioLLMs face significant challenges when processing long financial audio data. The SALMONN series models performed poorly on both ASR tasks. Upon reviewing the experimental logs, we observed that SALMONN frequently produced empty outputs or failed to understand prompt instructions. This suggests that SALMONN might be overfitting to specific audio data or tasks, lacks robust instruction-following capabilities, and exhibits a limited understanding of financial audio content.

In Figure~\ref{fig:results}, we find no significant difference in performance between Gemini-1.5 and Gemini-2.0. The open-source model Whisper-v3 achieves the best overall performance across all datasets. These results demonstrate robust capabilities even within long financial audio recognition. The Qwen-2 series, another open-source model, delivers comparable results to the closed-source Gemini models on short-audio ASR tasks but performs slightly worse on long-audio ASR. These findings demonstrate that open-source models can outperform their closed-source counterparts in financial audio scenarios, indicating that open-source AudioLLMs represent an effective, cost-efficient, and privacy-preserving solution for developing AI-driven financial products.
\vspace{-0.8em}

\subsection{Results on summarization task and ablation study}
\vspace{-0.7em}
For summarization, we evaluated Whisper-v3, Qwen2-Audio-7B-Instruct, Gemini-1.5-flash and GPT-4o-audio. The overall summarization quality depends on the ASR performance of AudioLLMs. Gemini-1.5-flash achieved the best ROUGE-L and BERTScore, while Whisper slightly outperformed Qwen2-Audio-7B-Instruct, showing strong general summarization ability without domain-specific tuning. We describe the detailed results in Appendix~\ref{sec:summary}

We also examined prompt robustness across ASR tasks. Whisper-v3 and Qwen2-Audio-7B-Instruct maintained stable performance under varied prompts, whereas Qwen2-Audio-7B was more sensitive, highlighting the importance of instruction tuning. Finally, our error analysis revealed that AudioLLMs often struggle with financial terminology and numerical information, underscoring the need for improved domain adaptation. We show the detailed analysis and visualization in Appendix~\ref{sec:robust}. We also conduct error analysis in Appendix~\ref{sec:error}.
\vspace{-0.7em}

\section{Conclusion and Research Outlook}
\vspace{-0.5em}
We propose \textsc{FinAudio}, an AudioLLM Benchmark for financial audio data. We constructed a total of 400h+ of audio data for three tasks and evaluated it on seven AudioLLMs. Our systematic evaluation revealed substantial challenges, notably the pronounced difficulty models face in accurately processing long-form financial audio and specialized terminology.  Future research directions for applying AudioLLMs in the financial industry include (1) extending the input context window length of AudioLLM, (2) enhancing the models' comprehension of numerical transcription and specialized financial terminology, and (3) improving their reasoning capabilities in financial contexts.

% \begin{ack}
% Use unnumbered first level headings for the acknowledgments. All acknowledgments
% go at the end of the paper before the list of references. Moreover, you are required to declare
% funding (financial activities supporting the submitted work) and competing interests (related financial activities outside the submitted work).
% More information about this disclosure can be found at: \url{https://neurips.cc/Conferences/2025/PaperInformation/FundingDisclosure}.

% Do {\bf not} include this section in the anonymized submission, only in the final paper. You can use the \texttt{ack} environment provided in the style file to automatically hide this section in the anonymized submission.
% \end{ack}

\medskip

\newpage

\bibliographystyle{unsrtnat}
\bibliography{main.bib}

@inproceedings{wu2023multimodal,
  title={{Multimodal large language models: A survey}},
  author={Wu, Jiayang and Gan, Wensheng and Chen, Zefeng and Wan, Shicheng and Philip, S Yu},
  booktitle={2023 IEEE International Conference on Big Data (BigData)},
  pages={2247--2256},
  year={2023},
  organization={IEEE}
}

@article{zhang2024mm,
  title={{MM-LLMs}: Recent advances in multimodal large language models},
  author={Zhang, Duzhen and Yu, Yahan and Dong, Jiahua and Li, Chenxing and Su, Dan and Chu, Chenhui and Yu, Dong},
  journal={arXiv preprint arXiv:2401.13601},
  year={2024}
}

@article{hu2024wavllm,
  title={{WavLLM}: Towards robust and adaptive speech large language model},
  author={Hu, Shujie and Zhou, Long and Liu, Shujie and Chen, Sanyuan and Meng, Lingwei and Hao, Hongkun and Pan, Jing and Liu, Xunying and Li, Jinyu and Sivasankaran, Sunit and others},
  journal={arXiv preprint arXiv:2404.00656},
  year={2024}
}

@article{chu2023qwen,
  title={{Qwen-Audio}: Advancing universal audio understanding via unified large-scale audio-language models},
  author={Chu, Yunfei and Xu, Jin and Zhou, Xiaohuan and Yang, Qian and Zhang, Shiliang and Yan, Zhijie and Zhou, Chang and Zhou, Jingren},
  journal={arXiv preprint arXiv:2311.07919},
  year={2023}
}

@article{chu2024qwen2,
  title={{Qwen2-Audio} technical report},
  author={Chu, Yunfei and Xu, Jin and Yang, Qian and Wei, Haojie and Wei, Xipin and Guo, Zhifang and Leng, Yichong and Lv, Yuanjun and He, Jinzheng and Lin, Junyang and others},
  journal={arXiv preprint arXiv:2407.10759},
  year={2024}
}

@inproceedings{radford2023robust,
  title={Robust speech recognition via large-scale weak supervision},
  author={Radford, Alec and Kim, Jong Wook and Xu, Tao and Brockman, Greg and McLeavey, Christine and Sutskever, Ilya},
  booktitle={International conference on machine learning},
  pages={28492--28518},
  year={2023},
  organization={PMLR}
}

@inproceedings{tangsalmonn,
  title={{SALMONN}: Towards Generic Hearing Abilities for Large Language Models},
  author={Tang, Changli and Yu, Wenyi and Sun, Guangzhi and Chen, Xianzhao and Tan, Tian and Li, Wei and Lu, Lu and Zejun, MA and Zhang, Chao},
  booktitle={The Twelfth International Conference on Learning Representations}
}

@inproceedings{yang2024air,
  title={{AIR-Bench}: Benchmarking Large Audio-Language Models via Generative Comprehension},
  author={Yang, Qian and Xu, Jin and Liu, Wenrui and Chu, Yunfei and Jiang, Ziyue and Zhou, Xiaohuan and Leng, Yichong and Lv, Yuanjun and Zhao, Zhou and Zhou, Chang and others},
  booktitle={Proceedings of the 62nd Annual Meeting of the Association for Computational Linguistics (Volume 1: Long Papers)},
  pages={1979--1998},
  year={2024}
}

@article{wang2024audiobench,
  title={{AudioBench}: A universal benchmark for audio large language models},
  author={Wang, Bin and Zou, Xunlong and Lin, Geyu and Sun, Shuo and Liu, Zhuohan and Zhang, Wenyu and Liu, Zhengyuan and Aw, AiTi and Chen, Nancy F},
  journal={arXiv preprint arXiv:2406.16020},
  year={2024}
}

@inproceedings{qin2019you,
  title={What you say and how you say it matters: Predicting stock volatility using verbal and vocal cues},
  author={Qin, Yu and Yang, Yi},
  booktitle={Proceedings of the 57th Annual Meeting of the Association for Computational Linguistics},
  pages={390--401},
  year={2019}
}

@inproceedings{yang2020html,
  title={Html: Hierarchical transformer-based multi-task learning for volatility prediction},
  author={Yang, Linyi and Ng, Tin Lok James and Smyth, Barry and Dong, Riuhai},
  booktitle={Proceedings of The Web Conference 2020},
  pages={441--451},
  year={2020}
}

@inproceedings{cao2024ecc,
  title={{ECC Analyzer}: Extracting Trading Signal from Earnings Conference Calls using Large Language Model for Stock Volatility Prediction},
  author={Cao, Yupeng and Chen, Zhi and Pei, Qingyun and Lee, Nathan and Subbalakshmi, KP and Ndiaye, Papa Momar},
  booktitle={Proceedings of the 5th ACM International Conference on AI in Finance},
  pages={257--265},
  year={2024}
}

@inproceedings{liu2023fingpt,
  title={{FinGPT}: Democratizing Internet-scale Data for Financial Large Language Models},
  author={Liu, Xiao-Yang and Wang, Guoxuan and Yang, Hongyang and Zha, Daochen},
  booktitle={NeurIPS 2023 Workshop on Instruction Tuning and Instruction Following}
}

@article{wu2023bloomberggpt,
  title={{BloombergGPT}: A large language model for finance},
  author={Wu, Shijie and Irsoy, Ozan and Lu, Steven and Dabravolski, Vadim and Dredze, Mark and Gehrmann, Sebastian and Kambadur, Prabhanjan and Rosenberg, David and Mann, Gideon},
  journal={arXiv preprint arXiv:2303.17564},
  year={2023}
}

@article{yang2023investlm,
  title={{InvestLM: A large language model for investment using financial domain instruction tuning}},
  author={Yang, Yi and Tang, Yixuan and Tam, Kar Yan},
  journal={arXiv preprint arXiv:2309.13064},
  year={2023}
}

@article{shah2022flue,
  title={When flue meets flang: Benchmarks and large pre-trained language model for financial domain},
  author={Shah, Raj Sanjay and Chawla, Kunal and Eidnani, Dheeraj and Shah, Agam and Du, Wendi and Chava, Sudheer and Raman, Natraj and Smiley, Charese and Chen, Jiaao and Yang, Diyi},
  journal={arXiv preprint arXiv:2211.00083},
  year={2022}
}

@article{xie2023pixiu,
  title={{PIXIU: A comprehensive benchmark, instruction dataset and large language model for finance}},
  author={Xie, Qianqian and Han, Weiguang and Zhang, Xiao and Lai, Yanzhao and Peng, Min and Lopez-Lira, Alejandro and Huang, Jimin},
  journal={Advances in Neural Information Processing Systems},
  volume={36},
  pages={33469--33484},
  year={2023}
}

@article{xie2025finben,
  title={Finben: A holistic financial benchmark for large language models},
  author={Xie, Qianqian and Han, Weiguang and Chen, Zhengyu and Xiang, Ruoyu and Zhang, Xiao and He, Yueru and Xiao, Mengxi and Li, Dong and Dai, Yongfu and Feng, Duanyu and others},
  journal={Advances in Neural Information Processing Systems},
  volume={37},
  pages={95716--95743},
  year={2025}
}

@article{xie2024open,
  title={Open-finllms: Open multimodal large language models for financial applications},
  author={Xie, Qianqian and Li, Dong and Xiao, Mengxi and Jiang, Zihao and Xiang, Ruoyu and Zhang, Xiao and Chen, Zhengyu and He, Yueru and Han, Weiguang and Yang, Yuzhe and others},
  journal={arXiv preprint arXiv:2408.11878},
  year={2024}
}

@inproceedings{bhatia2024fintral,
  title={FinTral: A Family of {GPT-4} Level Multimodal Financial Large Language Models},
  author={Bhatia, Gagan and Cavusoglu, Hasan and Abdul-Mageed, Muhammad and others},
  booktitle={Findings of the Association for Computational Linguistics ACL 2024},
  pages={13064--13087},
  year={2024}
}

@article{su2025audio,
  title={Audio-Language Models for Audio-Centric Tasks: A survey},
  author={Su, Yi and Bai, Jisheng and Xu, Qisheng and Xu, Kele and Dou, Yong},
  journal={arXiv preprint arXiv:2501.15177},
  year={2025}
}

@article{team2023gemini,
  title={Gemini: a family of highly capable multimodal models},
  author={Team, Gemini and Anil, Rohan and Borgeaud, Sebastian and Alayrac, Jean-Baptiste and Yu, Jiahui and Soricut, Radu and Schalkwyk, Johan and Dai, Andrew M and Hauth, Anja and Millican, Katie and others},
  journal={arXiv preprint arXiv:2312.11805},
  year={2023}
}

@inproceedings{huang2024audiogpt,
  title={{AudioGPT}: Understanding and generating speech, music, sound, and talking head},
  author={Huang, Rongjie and Li, Mingze and Yang, Dongchao and Shi, Jiatong and Chang, Xuankai and Ye, Zhenhui and Wu, Yuning and Hong, Zhiqing and Huang, Jiawei and Liu, Jinglin and others},
  booktitle={Proceedings of the AAAI Conference on Artificial Intelligence},
  volume={38},
  number={21},
  pages={23802--23804},
  year={2024}
}

@inproceedings{zhang2023speechgpt,
  title={SpeechGPT: Empowering Large Language Models with Intrinsic Cross-Modal Conversational Abilities},
  author={Zhang, Dong and Li, Shimin and Zhang, Xin and Zhan, Jun and Wang, Pengyu and Zhou, Yaqian and Qiu, Xipeng},
  booktitle={Findings of the Association for Computational Linguistics: EMNLP 2023},
  pages={15757--15773},
  year={2023}
}

@inproceedings{su2023pandagpt,
  title={{PandaGPT}: One Model To Instruction-Follow Them All},
  author={Su, Yixuan and Lan, Tian and Li, Huayang and Xu, Jialu and Wang, Yan and Cai, Deng},
  booktitle={Proceedings of the 1st Workshop on Taming Large Language Models: Controllability in the era of Interactive Assistants!},
  pages={11--23},
  year={2023}
}

@article{rubenstein2023audiopalm,
  title={{AudioPALM}: A large language model that can speak and listen},
  author={Rubenstein, Paul K and Asawaroengchai, Chulayuth and Nguyen, Duc Dung and Bapna, Ankur and Borsos, Zal{\'a}n and Quitry, F{\'e}lix de Chaumont and Chen, Peter and Badawy, Dalia El and Han, Wei and Kharitonov, Eugene and others},
  journal={arXiv preprint arXiv:2306.12925},
  year={2023}
}

@inproceedings{Yu2022DeepTF,
  title={Deep tensor factorization models of ﬁrst impressions},
  author={Yang Yu and Jordan W. Suchow},
  year={2022},
  url={https://api.semanticscholar.org/CorpusID:253065093}
}

@inproceedings{ang2022guided,
  title={Guided attention multimodal multitask financial forecasting with inter-company relationships and global and local news},
  author={Ang, Gary and Lim, Ee-Peng},
  booktitle={Proceedings of the 60th Annual Meeting of the Association for Computational Linguistics (Volume 1: Long Papers)},
  pages={6313--6326},
  year={2022}
}

@article{o2021spgispeech,
  title={{SPGISpeech: 5,000 hours of transcribed financial audio for fully formatted end-to-end speech recognition}},
  author={O'Neill, Patrick K and Lavrukhin, Vitaly and Majumdar, Somshubra and Noroozi, Vahid and Zhang, Yuekai and Kuchaiev, Oleksii and Balam, Jagadeesh and Dovzhenko, Yuliya and Freyberg, Keenan and Shulman, Michael D and others},
  journal={arXiv preprint arXiv:2104.02014},
  year={2021}
}

@article{del2021earnings,
  title={Earnings-21: A practical benchmark for {ASR} in the wild},
  author={Del Rio, Miguel and Delworth, Natalie and Westerman, Ryan and Huang, Michelle and Bhandari, Nishchal and Palakapilly, Joseph and McNamara, Quinten and Dong, Joshua and Zelasko, Piotr and Jett{\'e}, Miguel},
  journal={arXiv preprint arXiv:2104.11348},
  year={2021}
}

@article{del2022earnings,
  title={Earnings-22: A practical benchmark for accents in the wild},
  author={Del Rio, Miguel and Ha, Peter and McNamara, Quinten and Miller, Corey and Chandra, Shipra},
  journal={arXiv preprint arXiv:2203.15591},
  year={2022}
}

@inproceedings{mukherjee2022ectsum,
  title={{ECTSum}: A New Benchmark Dataset For Bullet Point Summarization of Long Earnings Call Transcripts},
  author={Mukherjee, Rajdeep and Bohra, Abhinav and Banerjee, Akash and Sharma, Soumya and Hegde, Manjunath and Shaikh, Afreen and Shrivastava, Shivani and Dasgupta, Koustuv and Ganguly, Niloy and Ghosh, Saptarshi and others},
  booktitle={Proceedings of the 2022 Conference on Empirical Methods in Natural Language Processing},
  pages={10893--10906},
  year={2022}
}

@inproceedings{lin2004rouge,
  title={{ROUGE}: A package for automatic evaluation of summaries},
  author={Lin, Chin-Yew},
  booktitle={Text summarization branches out},
  pages={74--81},
  year={2004}
}

@inproceedings{zhangbertscore,
  title={{BERTScore}: Evaluating Text Generation with BERT},
  author={Zhang, Tianyi and Kishore, Varsha and Wu, Felix and Weinberger, Kilian Q and Artzi, Yoav},
  booktitle={International Conference on Learning Representations}
}

%%%%%%%%%%%%%%%%%%%%%%%%%%%%%%%%%%%%%%%%%%%%%%%%%%%%%%%%%%%%

\appendix

\section{Related Work}
\subsection{Advancements in AudioLLM}
Among the modalities addressed by multimodal LLMs, audio stands out as particularly critical and challenging~\cite{su2025audio}. Several multimodal models, such as Gemini Pro~\cite{team2023gemini} and PandaGPT~\cite{su2023pandagpt}, have been adapted to process audio data. Prominent AudioLLMs include AudioGPT~\cite{huang2024audiogpt}, SpeechGPT~\cite{zhang2023speechgpt}, AudioPaLM~\cite{rubenstein2023audiopalm}, Qwen-Audio~\cite{chu2023qwen}, WavLLM~\cite{hu2024wavllm}, Whisper~\cite{radford2023robust}, Salmonn~\cite{tangsalmonn}, and Qwen2-Audio~\cite{chu2024qwen2}. Correspondingly, general-domain benchmarks like AirBench~\cite{yang2024air}, emphasizing conversational scenarios, and AudioBench~\cite{wang2024audiobench}, evaluating eight tasks across twenty datasets, have been developed to assess AudioLLMs. However, these benchmarks contain limited financial audio data, inadequate for a comprehensive evaluation in financial contexts. 

\begin{figure}[h]
\begin{center}
\centerline{\includegraphics[width=0.8\columnwidth]{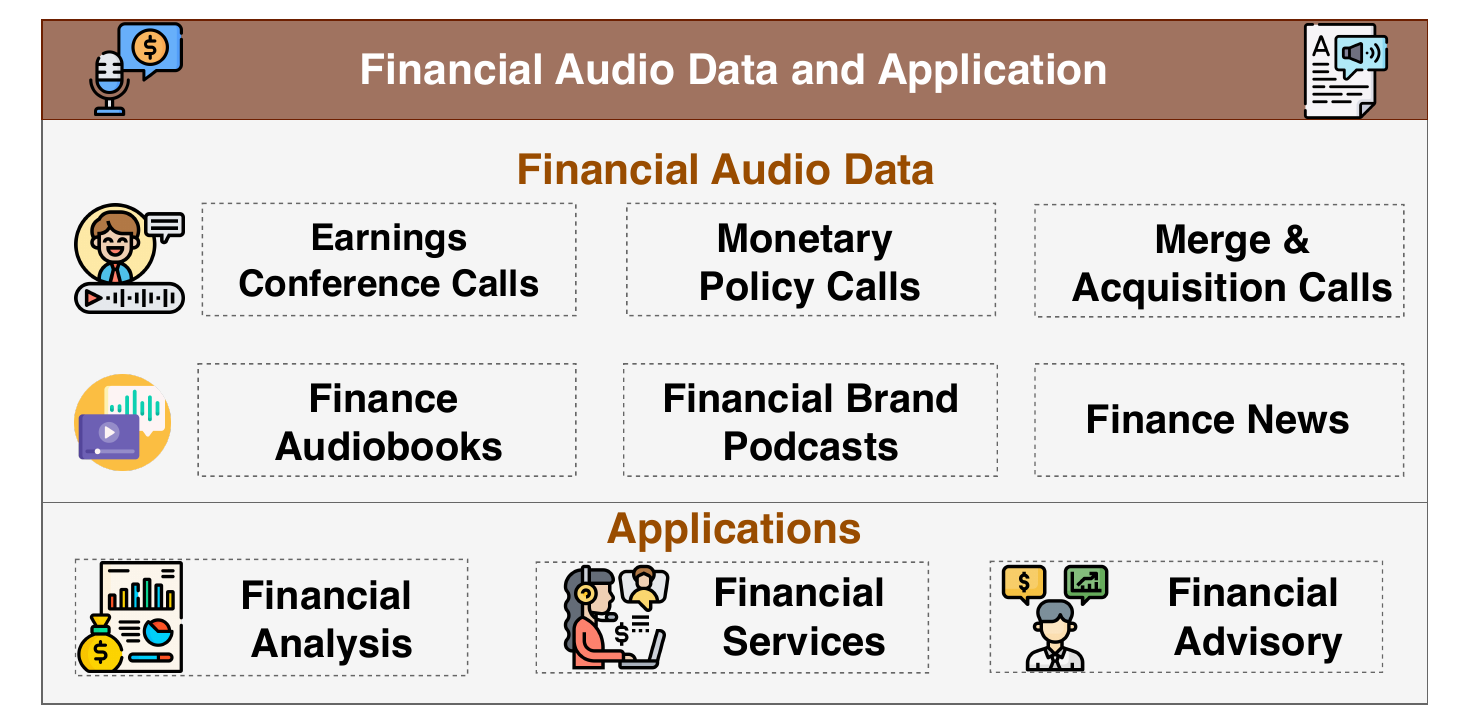}}
\caption{An overview of critical financial audio data types and applications.}% \textit{Note: WER refers to the term of Word Error Rate.}}
\label{fig:overview}
\end{center}
\end{figure}

%To address this gap, we introduce FinAudio, a holistic multimodal benchmark specifically designed for evaluating AudioLLMs in financial applications.
%Multimodal large language models have recently undergone rapid development. 
% Among the various modalities addressed in multimodal LLMs, audio stands out as one of the most critical and challenging domains~\cite{su2025audio}. Several multimodal LLMs, such as Gemini Pro~\cite{team2023gemini} and PandaGPT~\cite{su2023pandagpt}, have been adapted to handle audio data. 
% %AudioLLMs specifically developed from multimodal foundation models aim to enhance speech and audio understanding. 
% Notable AudioLLMs include AudioGPT~\cite{huang2024audiogpt}, SpeechGPT~\cite{zhang2023speechgpt}, AudioPaLM~\cite{rubenstein2023audiopalm}, Qwen-Audio~\cite{chu2023qwen}, WavLLM~\cite{hu2024wavllm}, Whisper~\cite{radford2023robust}, Salmonn~\cite{tangsalmonn}, Qwen2-Audio~\cite{chu2024qwen2}. Meanwhile, several benchmarks evaluate AudioLLMs in general audio contexts. For instance, AirBench~\cite{yang2024air} covers a wide range of audio—speech, natural sounds, and music—emphasizing performance in conversational scenarios, while AudioBench~\cite{wang2024audiobench} assesses four commonly used AudioLLMs across eight tasks and twenty datasets. However, these benchmarks include only limited financial audio data—insufficient for a comprehensive evaluation in complex financial contexts. Therefore, we present FinAudio as a holistic multimodal financial benchmark to fill the current research gap. 

\subsection{Progress in Multimodal Financial LLMs And Benchmark}
Recent advances in general-domain LLMs have accelerated the emergence of specialized financial LLMs, such as FinGPT~\cite{liu2023fingpt}, InvestLM~\cite{yang2023investlm}, and BloombergGPT~\cite{wu2023bloomberggpt}. The progression toward multimodal financial LLMs—including Open-FinLLMs~\cite{xie2024open} and FinTral~\cite{bhatia2024fintral}—further highlights efforts to address complex financial tasks involving charts and tabular data. These multimodal models overcome the limitations of traditional approaches~\cite{ang2022guided, bhatia2024fintral, Yu2022DeepTF} by more flexibly adapting to financial contexts. As financial LLMs become increasingly specialized, the demand for robust evaluation grows. In response, various benchmarks have been proposed.  FLUE~\cite{shah2022flue} introduced the first benchmark for financial NLP evaluation, covering tasks such as sentiment analysis, headline classification, named entity recognition, structure boundary detection, and question answering. PIXIU~\cite{xie2023pixiu} extended these evaluations into multimodal document analysis, while FinBen~\cite{xie2025finben} significantly expanded coverage, featuring 36 datasets across 24 financial tasks. Despite these advances, existing benchmarks remain predominantly text-focused, overlooking audio—an essential modality in financial applications.

%%%%%%%%%%%%%%%%%%%%%%%%%%%%%%%%%%%%%%%%%%%%%%%%%%%%%%%%%%%

\section{Dataset}
\label{sec:dataset}
\subsection{Short financial audio clip datasets:}
Two datasets provide financial audio clips: MDRM~\cite{qin2019you} and SPGISpeech~\cite{o2021spgispeech}, both derived from earnings conference call recordings. The original audio data was segmented at the sentence level and aligned with corresponding transcripts. Each dataset is initially split into training and test subsets; we utilize only the test sets for evaluation.

\begin{itemize}
    \item \textbf{MDRM~\cite{qin2019you}:} This is the first dataset to use multimodal financial data for stock volatility and movement prediction. MDRM includes both audio recordings and text transcripts from the 2017 earnings calls of 500 major companies listed on the S\&P 500 and traded on U.S. stock exchanges. There are a total of 280 companies with 572 earnings conference calls. The author further segmented earnings conference calls into individual sentences based on the text and conducted corresponding audio segments. There are a total of 88,829 audio clips. We split the dataset into training and test sets using a 75\%–25\% ratio, with the test portion utilized as the evaluation set for the benchmark. The MDRM-test set includes 22,208 audio clips totaling 87 hours.

    \item \textbf{SPGISpecch~\cite{o2021spgispeech}:} The dataset comprises 5,000 hours of earnings conference call recordings. The original audio is segmented by sentence and professionally transcribed into a structured format. Each audio segment is meticulously segmented by sentences, ranging from 5 to 15 seconds in duration. Partitioned initially into training and test sets, our analysis focuses solely on the test subset, comprising 39,341 clips and approximately 130 hours of audio.
\end{itemize}

\subsection{Long financial audio recording datasets:}
There are two other publicly available earnings conference call datasets containing raw audio recordings, which can be leveraged for the long financial audio ASR task:
\begin{itemize}
    \item \textbf{Earnings-21~\cite{del2021earnings}:} The dataset comprises 44 complete earnings conference call recordings from 2021 (approximately 39 hours). A distinguishing characteristic of Earnings-21 is its emphasis on named entities, financial jargon, and numerical data, which often pose difficulties for ASR models. The aligned transcripts are further enhanced with detailed formatting and entity tags, making it possible to conduct granular error analysis. 
     \item \textbf{Earnings-22~\cite{del2022earnings}:} This is a updated version of Earnings-21. It has 125 audio samples with a total of 120 hours.
Each audio recording includes a corresponding manual transcript. We use both datasets entirely for evaluation.
\end{itemize}

\subsection{FinAudioSum}
We introduce \textit{FinAudioSum}, a novel dataset created explicitly for long-form financial audio summarization tasks, building upon the foundation of the ECTSum dataset~\cite{mukherjee2022ectsum}. While ECTSum was initially designed for summarizing earnings calls based solely on textual transcripts, our contribution uniquely expands this resource into the audio modality, addressing a critical gap in multimodal financial summarization research. ECTSum comprises 2,425 earnings transcripts paired with expert-generated, telegram-style summaries. We obtain corresponding audio recordings for the ECTSum test set from \textsc{earningcast}\footnote{\url{https://earningscast.com/}}. To ensure uniqueness and avoid redundancy, we carefully eliminate any overlapping audio entries from Earnings-21 and Earnings-22 datasets, covering the years 2019–2022. The final \textit{FinAudioSum} dataset includes 64 recordings totaling 55 hours.

\begin{table}[ht]
\centering
\renewcommand{\arraystretch}{1.3} % row height
\begin{adjustbox}{max width=0.95\columnwidth}
\begin{tabular}{p{3cm} p{5cm} p{5cm} p{5cm}}
\toprule
\textbf{Task} & \textbf{Input} & \textbf{Instruction} & \textbf{Output} \\
\midrule
\textit{ASR for Short financial audio clip} & \includegraphics[width=0.025\textwidth]{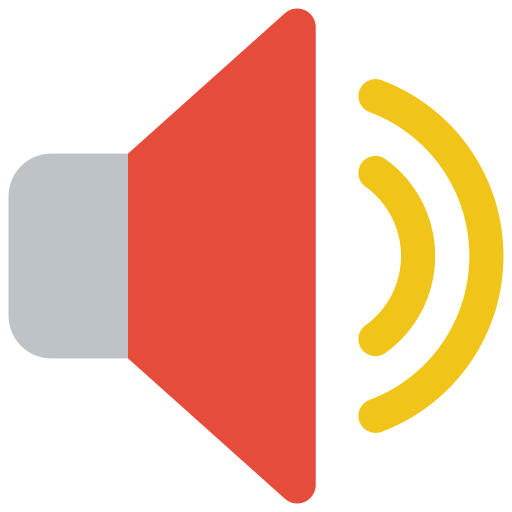} fast fashion has made it possible for everyone to have access to aesthetically thoughtful clothing (7s) & Convert the audio speech into a text transcript. & fast fashion has made it possible for everyone to have access to aesthetically thoughtful clothing \\
\midrule
\textit{ASR for Long financial audio recording} & \includegraphics[width=0.025\textwidth]{img/audio.png} Good morning, everyone, and welcome to the NextEra Energy Inc... (40min+) & Transcribe the spoken words into written form. & Good morning, everyone, and welcome to the NextEra Energy Inc...\\
\midrule
\textit{FinAudioSum} & \includegraphics[width=0.025\textwidth]{img/audio.png} Good morning, everyone, and welcome to the NextEra Energy Inc... (40min+) & Listen to the audio and provide the text
version. & In past three month, NextEra Energy Inc release...\\
\bottomrule
\end{tabular}
\end{adjustbox}
\caption{Examples of data in the \textsc{FinAudio} evaluation suite.}
\vspace{-2.5em}
\end{table}

\section{Model and Experiment Setup}
\label{setup}
We evaluate the performance of seven representative AudioLLMs on the \textsc{FinAudio} benchmark. The chosen models include five open-source models—Whisper~\cite{radford2023robust}, Qwen2-Audio-7B, Qwen2-Audio-7B-instruct~\cite{chu2024qwen2}, SALMONN-7B, and SALMONN-13B~\cite{tangsalmonn}—and three closed-source models, Gemini-1.5-flash and Gemini-2.0-flash~\cite{team2023gemini}, and GPT-4o-audio.  These models span various parameter sizes, open- and closed-source implementations, and include both base and instruction-tuned versions, enabling a comprehensive assessment of AudioLLMs for financial audio tasks. This comprehensive evaluation enables a thorough analysis of current AudioLLMs' strengths and limitations within financial audio contexts.

To ensure fairness, we standardized the prompt as: ``Convert the audio speech into a text transcript,''. For the summarization task, we employed GPT-4o (temperature=0) to generate summaries from the model-produced transcripts, ensuring a uniform summarization baseline. All open-source AudioLLM evaluations are conducted on two NVIDIA RTX A6000 GPUs, each equipped with 48GB of DRAM, necessitating approximately 200 hours per evaluation round. Additionally, to account for experimental variability and ensure the robustness of our assessments, each experiment was conducted three times, with the reported results representing the average performance.

\section{Results on summarization task}
\label{sec:summary}
\begin{table}[h]
\centering
\renewcommand{\arraystretch}{1.5}
\begin{adjustbox}{max width=0.95\columnwidth}
\small
\begin{tabular}{lcc|cc|cc|cc} % <-- added two more 'c' here
\toprule
\multirow{2}{*}{\textbf{Dataset}} 
 & \multicolumn{2}{c|}{\textbf{Whisper}} 
 & \multicolumn{2}{c|}{\textbf{Qwen2-Audio-7B-Instruct}} 
 & \multicolumn{2}{c|}{\textbf{Gemini-1.5-flash}} 
 & \multicolumn{2}{c}{\textbf{GPT-4o-audio}} \\ % <-- new model name
 & \textbf{Rouge-L} & \textbf{BertScore} 
 & \textbf{Rouge-L} & \textbf{BertScore} 
 & \textbf{Rouge-L} & \textbf{BertScore} 
 & \textbf{Rouge-L} & \textbf{BertScore} \\ % <-- new metric headers
\midrule
\large FinAudioSum 
 & \large 0.053 & \large 0.514 
 & \large 0.048 & \large 0.467 
 & \large \textbf{0.072} & \large \textbf{0.553} 
 & \large 0.063 & \large 0.508 \\ % <-- placeholder values
\bottomrule
\end{tabular}
\end{adjustbox}
\caption{Summarization performance using Rouge-L and BertScore.}
\label{tab:summarization}
\end{table}

Given the high WER in the ASR tasks, effective summarization cannot be reliably performed by models with excessive transcription errors. Therefore, we excluded the SALMONN model from the summarization evaluation. Balancing model performance and computational cost, we selected three models—Whisper-v3, Qwen2-Audio-7B-Instruct, and Gemini-1.5-flash for summarization experiments.

As Table~\ref{tab:summarization} shows, the summarization performance across all evaluated models is moderate. The overall summarization quality depends on the ASR performance of AudioLLMs. Gemini-1.5-flash achieves the highest scores in both ROUGE-L (0.072) and BERTScore (0.553), suggesting that Gemini's transcription accuracy and semantic fidelity during the ASR stage align more closely with the ground-truth labels. Whisper outperforms Qwen2-Audio-7B-Instruct (ROUGE-L: 0.053 vs. 0.048; BERTScore: 0.514 vs. 0.467), demonstrating its robust general summarization capability even without fine-tuning on financial domain data.

\section{Prompt Robustness Analysis}
\label{sec:robust}
The previous experiments indicated that different models respond differently to the prompt. Therefore, we further conducted a prompt robustness analysis on three AudioLLMs: Whisper-v3, Qwen2-Audio-7B, and Qwen2-Audio-7B-Instruct - across three ASR datasets. Following the approach from AudioBench~\cite{wang2024audiobench}, we constructed an instruction set comprising 10 distinct prompts to evaluate the robustness of AudioLLMs. These prompts convey similar instructions expressed in diverse formulations. The detailed prompts are listed as follows.

\begin{table}[h]
\centering
% \small
% \vspace{0.5em}
\begin{adjustbox}{max width=0.5\columnwidth}
\begin{tabular}{|p{1.2cm}|p{5.5cm}|}
\hline
\textbf{Number} & \textbf{Prompt Set} \\
\hline
1 & Convert the audio speech into a text transcript. \\
\hline
2 & Transcribe the spoken words into written form. \\
\hline
3 & Listen to the audio and provide the text version. \\
\hline
4 & Transform the speech into a text document. \\
\hline
5 & Capture the spoken language and convert it to text. \\
\hline
6 & Decode the audio and give me the written transcription. \\
\hline
7 & Recognize the verbal communication and transcribe it into text. \\
\hline
8 & Turn the speech input into a text transcription. \\
\hline
9 & Process the audio speech and provide the text output. \\
\hline
10 & Translate the spoken conversation into written text. \\
\hline
\end{tabular}
\end{adjustbox}
\caption{The evaluation instruction set for AudioLLMs}
\label{tab:prompt_table}
\end{table}

For each audio input, we randomly selected one prompt from this set to observe how variations in prompts affected model performance. This approach closely simulates real-world scenarios, where user commands often differ in wording yet convey the same intent.

\begin{wrapfigure}{r}{.50\textwidth}
    \centering
    \includegraphics[width=\linewidth]{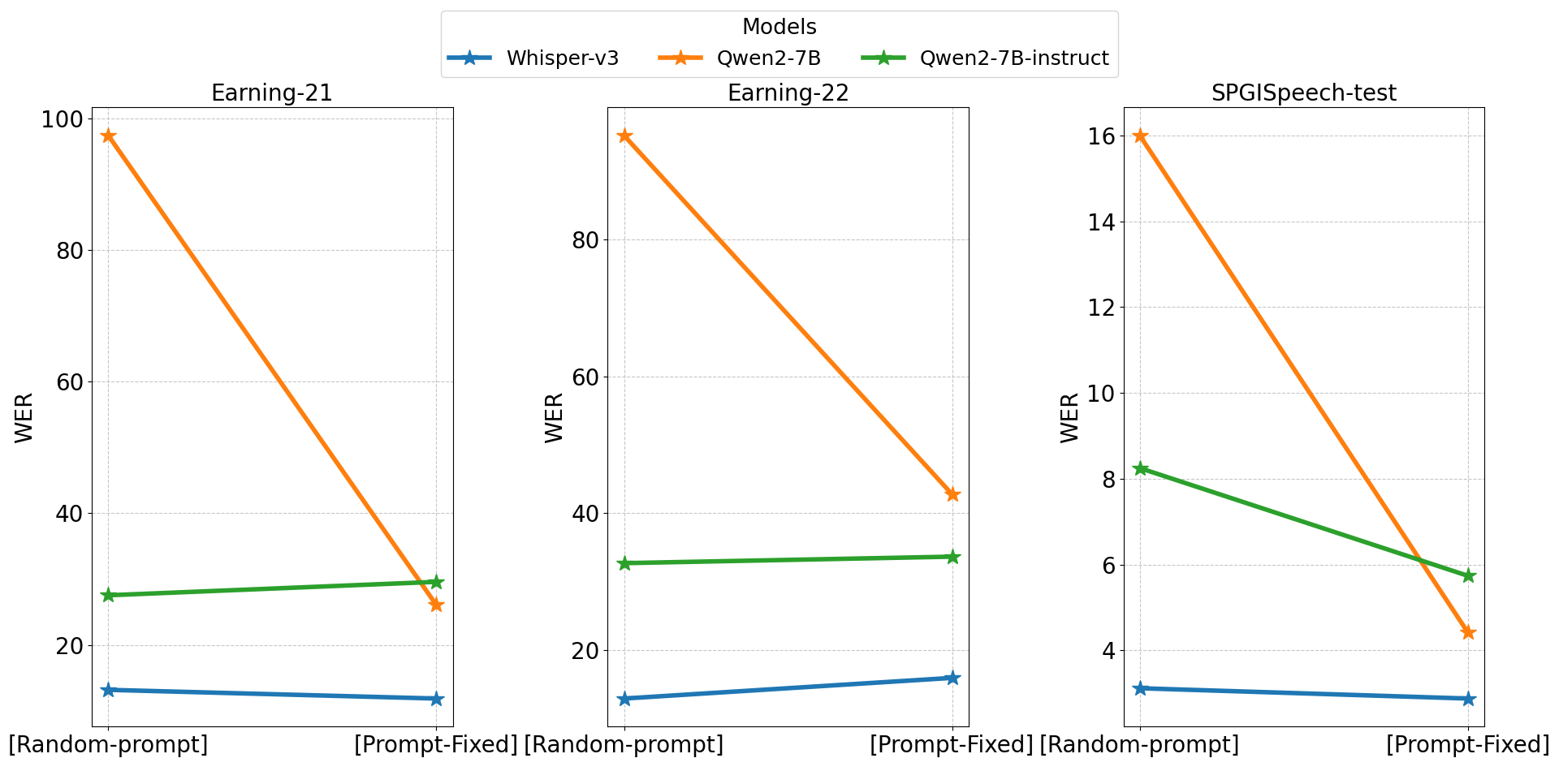}
    \caption{Prompt robustness analysis: comparison of WER between fixed-prompt and random-prompt trials.}
    \label{fig:prompt}
\end{wrapfigure}
We compare the WER in Figure~\ref{fig:prompt} using two prompt types: the random prompt described above and the fixed prompt from Section~\ref{setup}, used in the main results (Table~\ref{tab:comparison}). Results for Qwen2-7B-Instruct and Whisper-v3 remain consistent, indicating their robustness across prompts. However, the performance of Qwen2-7B notably declines under varied prompts, exhibiting higher WER. The substantial performance gap between Qwen2-7B and Qwen2-7B-Instruct indicates that this AudioLLM is sensitive to prompt variations, providing empirical support for the necessity of instruction-tuning in AudioLLM training.

\section{Error Analysis}
\label{sec:error}
The experimental results demonstrate that ASR performance significantly influences downstream tasks, such as summarization. Therefore, we take an error analysis. During the ASR tasks, we noticed that AudioLLMs sometimes misinterpret numerical information and specialized financial terms in speech. Such errors significantly impact the factual accuracy of AudioLLMs within financial contexts. Through detailed analysis of the outputs, we identified the following major error categories: \textbf{1) Financial terminology error:} Financial audio frequently contains specialized financial proper nouns, leading AudioLLMs to produce translation errors. For example, the company name ``NextEra Energy" was incorrectly transcribed as ``Era Energy." \textbf{2) Numerical information error:} Financial audio typically includes extensive numerical information, such as monetary amounts. During transcription, AudioLLMs frequently encounter issues like digit inconsistencies or missing monetary units.

This analysis will guide future model training by prioritizing improvements in numerical recognition and enhancing the model's ability to accurately transcribe financial terminology.

\end{document}